\documentclass[letterpaper]{article} 
\usepackage{aaai2027}  
\usepackage[hyphens]{url}  
\usepackage{graphicx} 
\usepackage{natbib}  
\usepackage{caption} 
\usepackage{algorithm}
\usepackage{algorithmic}

\usepackage{newfloat}
\usepackage{listings}
\DeclareCaptionStyle{ruled}{labelfont=normalfont,labelsep=colon,strut=off} 
\floatstyle{ruled}
\newfloat{listing}{tb}{lst}{}
\floatname{listing}{Listing}

\usepackage{booktabs}

\title{Physics-Constrained Soft Actor-Critic for Simulator-in-the-Loop Petroleum Reservoir History Matching}
\author{
    Nam-Phong Huu Nguyen\textsuperscript{\rm 1},
    Duy-Dong Nguyen\textsuperscript{\rm 1},
    Tho Quan\textsuperscript{\rm 1},
}
\affiliations{
    \textsuperscript{\rm 1}Ho Chi Minh City University of Technology (HCMUT), VNU-HCM, Ho Chi Minh City, Vietnam\\

}

\begin{document}
\raggedbottom

\maketitle

\begin{abstract}
Many scientific calibration problems expose only an expensive executable simulator, making gradients unavailable and large-scale training-data generation impractical. We study petroleum reservoir history matching as an instance of this broader AI problem and formulate it as physics-constrained, simulator-in-the-loop policy search. Our method wraps the CMG IMEX full-physics simulator as a Gymnasium environment and uses Soft Actor-Critic (SAC) to learn a stochastic proposal distribution over continuous porosity, directional-permeability, and well-skin parameters. Each interaction generates and executes a reservoir case, aligns simulated and observed production responses, and returns a reward that combines multi-response mismatch with penalties for physically invalid properties. Off-policy replay reuses costly simulator feedback, while maximum-entropy learning preserves exploration. Unlike forward-surrogate approaches, the policy learns where to evaluate rather than learning to replace the simulator; every retained candidate is validated by IMEX. Under a 200-call budget on PUNQ-S3, the best valid candidate achieves category-macro NMSE $0.0285$, $R^2=0.9324$, and a bounded match score of $97.23\%$. Well-level analysis further exposes localized water-rate failures hidden by pooled metrics. These results establish a full-physics proof of concept for reinforcement-learning-based calibration of expensive scientific simulators.
\end{abstract}


\section{Introduction}

Reservoir forecasts depend on geological, flow, and well parameters that are only indirectly observed. History matching estimates these parameters by minimizing disagreement between simulated and historical production responses~\cite{oliver2011review,liu2023automatic}. The resulting objective is nonlinear and non-unique, and commercial simulators expose neither inexpensive evaluations nor useful gradients.

Existing automation follows two main AI strategies. Surrogate-based methods learn a fast approximation to the simulator, but require representative training simulations and final high-fidelity validation~\cite{badawi2025neural,aslam2025dnn,liu2025automated}. Reinforcement-learning methods instead learn how to generate calibration candidates from interaction data~\cite{li2021rlhistory,alolayan2023parallel}. We pursue the latter view while retaining CMG IMEX as the authoritative environment: SAC learns a stochastic proposal policy from replayed simulator evaluations, and the reward encodes both data fit and physical admissibility.

Overall, the contributions of our research can be articulated as follows:
\begin{itemize}
    \item We formulate history matching as physics-constrained, simulator-in-the-loop policy search and implement an end-to-end Gymnasium environment around CMG IMEX.
    \item We design a continuous, physically interpretable action space and a reward that combines multi-response production mismatch with penalties for inadmissible reservoir properties.
    \item We evaluate SAC on PUNQ-S3 with full-physics calls and report aggregate and well-response metrics, revealing failure modes that pooled scores conceal.
\end{itemize}
\section{Related Work}

\subsection{Simulation-Based History Matching}

Classical history matching combines parameterization, mismatch design, derivative-free optimization, and uncertainty analysis~\cite{oliver2011review,abdulrazzaq2023review,liu2023automatic}. EnKF, ES-MDA, and iterative smoothers update ensembles from sample covariances and quantify posterior uncertainty, but require repeated simulation of many realizations~\cite{aanonsen2009enkf,emerick2013esmda,evensen2019ies,chen2013lmies}. Differential Evolution and design-of-experiments methods provide gradient-free alternatives~\cite{storn1997de,li2019bestpractices}; GP-VARS reduces their cost through Gaussian-process proxies and sensitivity analysis~\cite{rana2018gpvars}.

\subsection{Reinforcement Learning as Black-Box Policy Search}

Learning-to-optimize replaces a fixed candidate-generation rule with a policy learned from evaluations~\cite{tang2024learn}. SAC is an off-policy maximum-entropy actor--critic method: replay improves data reuse, while a stochastic actor balances reward maximization and exploration in continuous action spaces~\cite{haarnoja2018sac,wang2024drlsurvey}. Prior reservoir work applied DQN and DDPG to history matching with a fast-marching simulator and later explored parallel RL search for multiple solutions~\cite{li2021rlhistory,alolayan2023parallel}. Our setting differs by coupling SAC to a commercial full-physics simulator and by explicitly penalizing invalid physical properties.

\subsection{Learned Reservoir Models}

Deep generative and feature-based parameterizations have been combined with ensemble smoothing~\cite{canchumuni2019generative,kim2020esnn,razak2020cnn}. Recurrent proxies, neural operators, and reduced-order models further reduce online simulation cost~\cite{tang2020surrogate,ma2022hybrid,zhang2023lstm,badawi2025neural,abdulkareem2026e2co}; graph neural networks and Transformers have also been used for interwell-connectivity inversion~\cite{liu2025automated}. These approaches can accelerate inference but depend on representative training data and full-physics validation~\cite{cui2025multifidelity,motaei2026proxyreview}. In contrast, our policy learns a search distribution, not a forward surrogate: every candidate used for model selection is evaluated by CMG IMEX.

\section{Methodology}

\begin{figure*}[t]
    \centering
    \includegraphics[width=\textwidth]{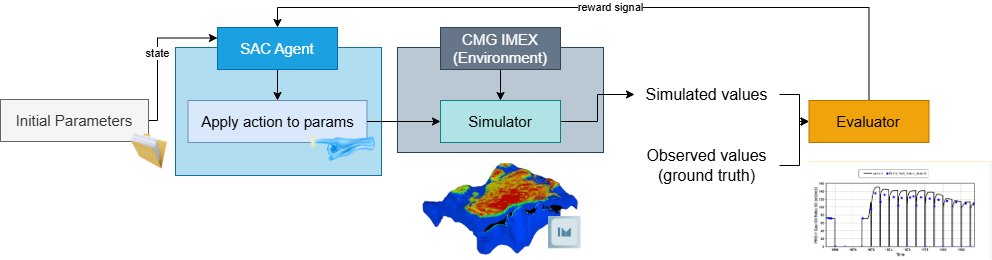}
    \caption{Simulator-in-the-loop history-matching workflow. The SAC agent transforms the current state into continuous parameter updates, CMG IMEX evaluates the resulting reservoir case, and the evaluator compares simulated and observed production responses to return the reward signal.}
    \label{fig:sac_imex_workflow}
\end{figure*}

We formulate history matching as constrained black-box policy search. At interaction $t$, a stochastic policy proposes continuous calibration parameters, CMG IMEX evaluates the resulting reservoir model, and the environment returns a scalar reward measuring production agreement and physical admissibility. This view exposes the simulator through an RL interface without assuming differentiability or learning a replacement simulator, as overviewed in Figure~\ref{fig:sac_imex_workflow}.

\begin{algorithm}[t]
\caption{Simulator-in-the-Loop Soft Actor-Critic for Reservoir History Matching}
\label{alg:sac_history_matching}
\footnotesize
\begin{algorithmic}[1]
\REQUIRE Base model $\mathcal{M}_0$, history $\mathcal{H}^{obs}$, bounds $(\mathbf{\ell},\mathbf{u})$, simulator $\mathcal{S}$, tolerance $\tau$, interactions $T$
\ENSURE Best matched parameters $\mathbf{\theta}^{\star}$
\STATE Initialize SAC networks and replay buffer $\mathcal{D}$; set $\mathbf{a}_0=\mathbf{0}$, base-model error $E_0$, and $E^{\star}=\infty$
\STATE Define $\mathbf{\theta}=[m_{\phi},m_{k_i},m_{k_j},m_{k_k},s_{\mathrm{skin}}]$
\FOR{$t=1,\ldots,T$}
    \STATE $\mathbf{o}_t\leftarrow[\mathbf{a}_{t-1},E_{t-1}]$; sample $\mathbf{a}_t\sim\pi_{\phi}(\cdot\mid\mathbf{o}_t)$ in $[-1,1]^5$
    \STATE $\mathbf{\theta}_t\leftarrow\mathbf{\ell}+\frac{\mathbf{a}_t+\mathbf{1}}{2}\odot(\mathbf{u}-\mathbf{\ell})$
    \STATE Apply $\mathbf{\theta}_t$ to $\mathcal{M}_0$ and run IMEX: $\mathcal{Y}^{sim}_t\leftarrow\mathcal{S}(\mathcal{M}_t)$
    \STATE Match simulated and observed responses by well and nearest time within $\tau$
    \STATE Set $\mathcal{Q}=\{\mathrm{BHP},\mathrm{Oil},\mathrm{Water},\mathrm{Gas}\}$
    \STATE $E_t\leftarrow\frac{\sum_{q\in\mathcal{Q}}w_q\,\ell_q(\mathbf{\theta}_t)}{\sum_{q\in\mathcal{Q}}w_q}$
    \STATE Compute $P_t\leftarrow P_{\mathrm{phys}}(\mathbf{\theta}_t)$ and set $r_t\leftarrow-E_t-P_t$
    \STATE Set $\mathbf{o}_{t+1}\leftarrow[\mathbf{a}_t,E_t]$ and store the terminal transition in $\mathcal{D}$
    \STATE Update the critics, policy, and entropy temperature from replay mini-batches
    \IF{$P_t=0$ and $E_t<E^{\star}$}
        \STATE Set $E^{\star}\leftarrow E_t$ and $\mathbf{\theta}^{\star}\leftarrow\mathbf{\theta}_t$; archive $\mathcal{M}_t$
    \ENDIF
\ENDFOR
\RETURN $\mathbf{\theta}^{\star}$
\end{algorithmic}
\end{algorithm}

\subsection{Problem Formulation}

Let $\mathbf{m}$ denote the vector of uncertain reservoir parameters to be calibrated. In this study, we consider a compact parameterization:
\[
\mathbf{m} =
\left[
m_{\phi},
m_{k_i},
m_{k_j},
m_{k_k},
s
\right],
\]
where $m_{\phi}$ is a global porosity multiplier, $m_{k_i}$, $m_{k_j}$, and $m_{k_k}$ are permeability multipliers along the $I$, $J$, and $K$ directions, and $s$ is a global well skin factor. These variables are selected because they directly affect storage capacity, transmissibility, and well productivity, which are key factors controlling production behavior.

The parameters are constrained to predefined physical ranges:
\[
\begin{array}{rcl}
m_{\phi} & \in & [0.80, 1.20],\\
m_{k_i},m_{k_j},m_{k_k} & \in & [0.30, 3.00],\\
s & \in & [-5.0, 15.0].
\end{array}
\]
The learning agent operates in a normalized action space,
\[
\mathbf{a} \in [-1,1]^5,
\]
and each action component is linearly mapped to its corresponding parameter range:
\[
m_i = l_i + \frac{a_i + 1}{2}(u_i - l_i),
\]
where $l_i$ and $u_i$ are the lower and upper bounds of the $i$-th calibration variable. This mapping gives SAC a standardized continuous action space while keeping proposals within the prescribed control ranges; the separate physics penalty handles invalid properties that may still arise after modifying the reservoir grid.

\subsection{Simulator-in-the-Loop Environment}

The environment couples the learning agent with the CMG IMEX reservoir simulator. For every action proposed by the agent, a new reservoir case is generated by modifying the base PUNQ-S3 model~\cite{floris2001punq,rana2018gpvars,alguliyev2022history}. The porosity and directional permeability fields are scaled according to the selected multipliers, and the well skin factor is updated consistently across well completions. The modified case is then executed using IMEX, and the simulator output is parsed to obtain well-level production responses.

The observed history is extracted from the field history file, while the simulated responses are extracted from the simulator output. We compare four production signals:
\[
\{P_{bhp}, Q_o, Q_w, Q_g\},
\]
corresponding to bottom-hole pressure, oil rate, water rate, and gas rate. Since reporting times may not coincide, records are matched by well and nearest time within a fixed tolerance. Each response uses the records for which that simulated--observed pair is available, permitting different sample counts across responses.

\subsection{Objective and Reward Design}

Because the production signals have different units and scales, the training objective uses a range-normalized per-response loss:
\[
\ell_q(\mathbf{m}) =
\frac{1}{N_q}
\sum_{n=1}^{N_q}
\left(
\frac{y^{sim}_{q,n} - y^{obs}_{q,n}}
{\max(y^{obs}_{q}) - \min(y^{obs}_{q})+\epsilon}
\right)^2.
\]
Here, $y^{sim}_{q,n}$ and $y^{obs}_{q,n}$ are matched simulated and observed values, and $\epsilon>0$ prevents division by zero. We denote this loss by $\ell_q$ to distinguish the reward objective from the evaluation NMSE defined later.
The total mismatch is computed as a weighted average over all available signals:
\[
E(\mathbf{m}) =
\frac{
\sum_q w_q \ell_q(\mathbf{m})
}{
\sum_q w_q
},
\]
where $w_q$ is the response weight. In the current implementation, all four signals receive equal weight.

The reinforcement learning reward combines the production mismatch with a penalty for physically invalid reservoir properties:
\[
r(\mathbf{m}) = -E(\mathbf{m}) - P_{\mathrm{phys}}(\mathbf{m}),
\]
where $P_{\mathrm{phys}}(\mathbf{m})\geq 0$ is the implemented penalty for violations of the prescribed physical-property constraints and $P_{\mathrm{phys}}(\mathbf{m})=0$ for physically valid candidates. Thus, maximizing the expected reward simultaneously reduces the history-matching error and discourages physically inadmissible reservoir configurations. Failed simulations, invalid outputs, or cases without sufficient matched data are also assigned a large penalty, discouraging the agent from selecting unstable or uninformative parameter configurations.

\subsection{Learning Procedure}

We use SAC to learn a stochastic proposal policy over the calibration space~\cite{haarnoja2018sac}. The policy receives a compact search-memory observation containing the previous action and mismatch:
\[
\mathbf{o}_t =
\left[
\mathbf{a}_{t-1},
E_{t-1}
\right].
\]
Each episode contains one simulator evaluation, so the formulation is bandit-like rather than a conventional long-horizon control task. SAC maximizes the maximum-entropy objective
\[
J(\pi)=\mathrm{E}_{\mathbf{a}\sim\pi(\cdot\mid\mathbf{o})}
\left[r(\mathbf{a})+\alpha\mathcal{H}\!\left(\pi(\cdot\mid\mathbf{o})\right)\right],
\]
where $\alpha$ controls exploration. Although transitions are terminal, off-policy replay reuses past simulator calls across actor and critic updates, and each reset observation carries a minimal summary of the preceding interaction.

For terminal transitions, the critic target is the observed immediate reward. The twin critics and stochastic actor are therefore trained with
\[
\mathcal{L}_{Q}(\psi_j)=
\mathrm{E}_{(\mathbf{o},\mathbf{a},r)\sim\mathcal{D}}
\left[\left(Q_{\psi_j}(\mathbf{o},\mathbf{a})-r\right)^2\right],
\quad j\in\{1,2\},
\]
and
\[
\mathcal{L}_{\pi}(\phi)=
\mathrm{E}\!\left[
\alpha\log\pi_{\phi}(\mathbf{a}\mid\mathbf{o})
-\min_j Q_{\psi_j}(\mathbf{o},\mathbf{a})
\right].
\]
In this setting, the critics approximate the penalized simulator-return landscape from replayed evaluations, while the actor shifts probability toward high-reward regions without immediately discarding exploratory actions. This interpretation makes explicit that SAC is used as a learned black-box search distribution rather than as a controller of reservoir dynamics.

Each candidate is evaluated by a full simulator run, and the best valid reservoir case is retained with its parameters and matching statistics.

Algorithm~\ref{alg:sac_history_matching} summarizes the complete simulator-in-the-loop optimization workflow.

\section{Experiments}
\subsection{Experiment Setup}
Experiments were performed on the PUNQ-S3 reservoir benchmark~\cite{floris2001punq,rana2018gpvars,alguliyev2022history} using CMG IMEX 2021.10 as the forward simulator. The input data consisted of the base CMG reservoir model and the corresponding observed field-history file. Each simulation case was executed in an isolated run directory, with generated input files, simulator outputs, matched production data, and logs stored under the experiment runs folder.

The uncertain parameters considered in the experiments were porosity multiplier, permeability multipliers in the I, J, and K directions, and a global producer skin factor. Their ranges were fixed as \([0.80,1.20]\), \([0.30,3.00]\), \([0.30,3.00]\), \([0.30,3.00]\), and \([-5.00,15.00]\), respectively. The production responses used for evaluation were bottom-hole pressure, oil rate, water rate, and gas rate, matched between simulated and observed records using a 30-day time tolerance.

Experiments were implemented in Python using a Gymnasium-compatible environment and Stable-Baselines3. SAC used an MLP policy, learning rate \(3\times10^{-4}\), replay capacity of 10,000 transitions, batch size 64, and 200 simulator interactions. We report one training run and retain the best physically valid candidate encountered. The workflow logs parameters, simulator status, reward components, matched-point coverage, and output paths. Published DE, GP-VARS, EnKF, deep ES-MDA, and FNO results are used only as contextual references because their objectives and simulation budgets differ.

Each interaction corresponds to one generated IMEX case and one stored replay transition; no learned surrogate or surrogate-training dataset is used. Penalized candidates remain informative to the policy through their rewards but are ineligible for best-model retention. Among candidates with zero physics penalty and valid simulator output, selection uses the production mismatch $E(\mathbf{m})$. This separates physical feasibility, learning feedback, and terminal model selection.

\subsection{Evaluation Protocol}

We evaluate the proposed RL-SAC history-matching framework on the PUNQ-S3 reservoir benchmark~\cite{floris2001punq,alguliyev2022history,rana2018gpvars}. The primary production responses include oil production rate, water production rate, gas production rate, and bottom-hole pressure (BHP) for six production wells: PRO-1, PRO-4, PRO-5, PRO-11, PRO-12, and PRO-15. We additionally derive the gas--oil ratio (GOR) and water cut (WC) from the matched oil, gas, and water records. At timestamps with nonzero denominators, the observed and simulated ratios are

\begin{equation}
\mathrm{GOR}^{\mathrm{obs}}_i =
\frac{G^{\mathrm{obs}}_i}{O^{\mathrm{obs}}_i},
\qquad
\mathrm{GOR}^{\mathrm{sim}}_i =
\frac{G^{\mathrm{sim}}_i}{O^{\mathrm{sim}}_i},
\end{equation}

\begin{equation}
\mathrm{WC}^{\mathrm{obs}}_i =
\frac{W^{\mathrm{obs}}_i}{O^{\mathrm{obs}}_i + W^{\mathrm{obs}}_i},
\qquad
\mathrm{WC}^{\mathrm{sim}}_i =
\frac{W^{\mathrm{sim}}_i}{O^{\mathrm{sim}}_i + W^{\mathrm{sim}}_i}.
\end{equation}

Records with a zero oil-rate denominator for GOR or a zero total-liquid denominator for WC are excluded from the corresponding ratio metric. NMSE, $R^2$, and $C_{\mathrm{match}}$ are then computed between the observed and simulated ratio sequences using the same definitions as the primary responses.

For reporting, we use an energy-normalized squared error (denoted NMSE to follow the reservoir literature) and the coefficient of determination ($R^2$):

\begin{equation} \mathrm{NMSE} = \frac{ \sum_{i=1}^{N} \left(y_i-\hat{y}_i\right)^2 }{ \sum_{i=1}^{N} y_i^2 + \epsilon }, \end{equation}

\begin{equation} R^2 = 1- \frac{ \sum_{i=1}^{N} \left(y_i-\hat{y}_i\right)^2 }{ \sum_{i=1}^{N} \left(y_i-\bar{y}\right)^2 + \epsilon }, \end{equation}

where $y_i$ and $\hat{y}_i$ denote the observed and simulated responses, respectively. Under this definition, $R^2$ compares the residual error with the error of predicting the observed mean~\cite{chicco2021r2,kvalseth1985r2}. To express terminal match quality on a bounded percentage scale, we define

\begin{equation} C_{\mathrm{match}} = \frac{100}{1+\mathrm{NMSE}}. \end{equation}

The bounded match score satisfies $0<C_{\mathrm{match}}\leq100$: an exact match gives $100\%$, and the score approaches zero as NMSE increases. It summarizes terminal fit quality and is not a measure of optimization convergence over simulator calls.




\subsection{Results}
\label{sec:results}

\subsubsection{Overall History-Matching Accuracy}

We evaluate the proposed RL-SAC framework on six production wells of the PUNQ-S3 reservoir, namely PRO-1, PRO-4, PRO-5, PRO-11, PRO-12, and PRO-15. The evaluated dynamic responses comprise oil production rate, water production rate, gas production rate, and bottom-hole pressure (BHP). Gas--oil ratio (GOR) and water cut are additionally evaluated as derived production responses.

Table~\ref{tab:category_results} summarizes the cleaned evaluation data. The best candidate achieves a category-macro NMSE of $0.0285$, category-macro $R^2$ of $0.9324$, and bounded match score of $97.23\%$, indicating strong aggregate agreement with field history.

\begin{table}[t]
    \centering
    \caption{Category-level performance on PUNQ-S3. $C_{\mathrm{match}}$ is the bounded match score. GOR and water cut are diagnostic responses excluded from the category macro.}
    \label{tab:category_results}
    \footnotesize
    \begin{tabular}{@{}lcccc@{}}
        \hline
        Response & $N$ & NMSE $\downarrow$ & $R^2$ $\uparrow$ & $C_{\mathrm{match}}$ (\%) $\uparrow$ \\
        \hline
        Oil rate   & 180 & 0.0278 & 0.9287 & 97.30 \\
        Water rate & 180 & 0.0509 & 0.9423 & 95.16 \\
        Gas rate   & 180 & 0.0334 & 0.9207 & 96.77 \\
        BHP         & 174 & 0.0018 & 0.9378 & 99.82 \\
        \hline
        GOR         & 114 & 0.0079 & 0.8977 & 99.21 \\
        Water cut   & 114 & 0.0176 & 0.9780 & 98.27 \\
        \hline
        Category macro & -- & 0.0285 & 0.9324 & 97.23 \\
        \hline
    \end{tabular}
\end{table}

\begin{table*}[t]
    \centering
    \caption{Well-level match score $C_{\mathrm{match}}$ (\%). Overall values use pooled category NMSE rather than the arithmetic mean of well scores. Ratios exclude zero denominators.}
    \label{tab:well_category_results}
    \footnotesize
    \renewcommand{\arraystretch}{1.15}
    \begin{tabular}{@{}lccccccc@{}}
        \hline
        Response & PRO-1 & PRO-4 & PRO-5 & PRO-11 & PRO-12 & PRO-15 & Overall \\
        \hline
        Oil rate   & 97.68 & 97.33 & 97.68 & 96.73 & 97.68 & 96.37 & 97.30 \\
        Water rate & 13.31 & 96.41 & 92.68 & 93.77 & 96.61 & 56.87 & 95.16 \\
        Gas rate   & 97.74 & 95.38 & 97.15 & 96.00 & 97.14 & 96.13 & 96.77 \\
        BHP        & 99.78 & 99.85 & 99.84 & 99.91 & 99.73 & 99.88 & 99.82 \\
        \hline
        GOR        & 98.83 & 97.83 & 99.88 & 99.90 & 99.91 & 99.58 & 99.21 \\
        Water cut  & 13.37 & 97.46 & 93.50 & 99.16 & 96.23 & 57.90 & 98.27 \\
        \hline
    \end{tabular}
\end{table*}

Among the four primary responses, BHP has the lowest NMSE ($0.0018$) and a match score of $99.82\%$. Oil and gas obtain scores of $97.30\%$ and $96.77\%$, respectively. Water has the largest category NMSE, although its pooled $R^2$ remains $0.9423$.

For diagnostic responses, GOR obtains NMSE $0.0079$, $R^2=0.8977$, and $C_{\mathrm{match}}=99.21\%$; water cut obtains NMSE $0.0176$, $R^2=0.9780$, and $C_{\mathrm{match}}=98.27\%$. Because both are derived from primary rates, they are excluded from the macro score.

The high aggregated water-production score, however, does not imply that all individual wells are matched equally well. We therefore examine the results at the well-response level.

\subsubsection{Well-Level Match Quality}

Table~\ref{tab:well_category_results} summarizes $C_{\mathrm{match}}$ by well and response. Oil and gas are consistently matched across all six wells, whereas water shows substantially greater variation.

The main mismatch occurs in the water-production response of PRO-1, for which NMSE increases to $6.5124$, $R^2$ becomes $-7.0474$, and $C_{\mathrm{match}}$ decreases to $13.31\%$. A negative $R^2$ indicates that the matched trajectory performs worse than predicting the mean observed response~\cite{chicco2021r2,kvalseth1985r2}. Water production at PRO-15 is also more difficult to reproduce, with NMSE $=0.7583$, $R^2=0.1998$, and $C_{\mathrm{match}}=56.87\%$.

These two cases explain why the category-level water NMSE is larger than the remaining responses. They also show that a strong field-level or category-level metric can conceal localized mismatch. In contrast, the oil and gas responses remain stable across all evaluated wells.

GOR remains consistently matched across the six wells, with well-level scores above $97.8\%$. Water cut inherits the localized water-rate mismatch at PRO-1 and PRO-15, while the remaining wells achieve scores above $93\%$.

The relatively low BHP $R^2$ at PRO-5 and PRO-12, despite their small NMSE values, can be attributed partly to the low variance of the corresponding observed pressure trajectories. Because the total sum of squares scales with the observed variance, modest absolute deviations can substantially reduce $R^2$ for nearly constant responses~\cite{chicco2021r2,kvalseth1985r2}.

\begin{figure*}[t]
    \centering
    \setlength{\tabcolsep}{2pt}
    \renewcommand{\arraystretch}{1.0}
    \begin{tabular}{cc}
        \includegraphics[
            width=0.315\textwidth,
            height=0.205\textheight,
            keepaspectratio
        ]{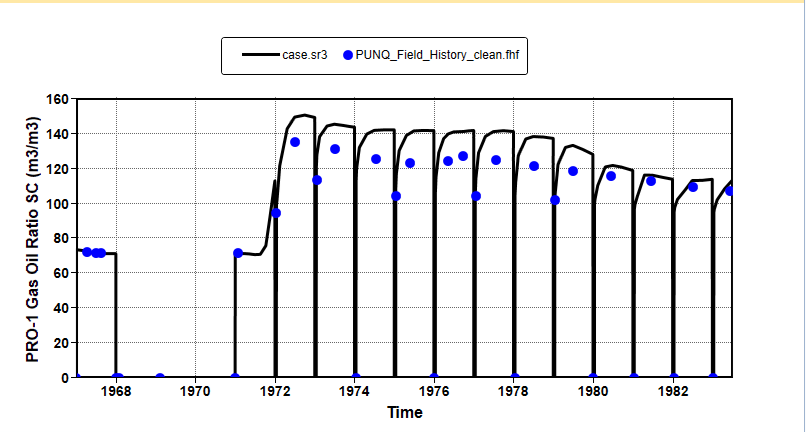}
        &
        \includegraphics[
            width=0.315\textwidth,
            height=0.205\textheight,
            keepaspectratio
        ]{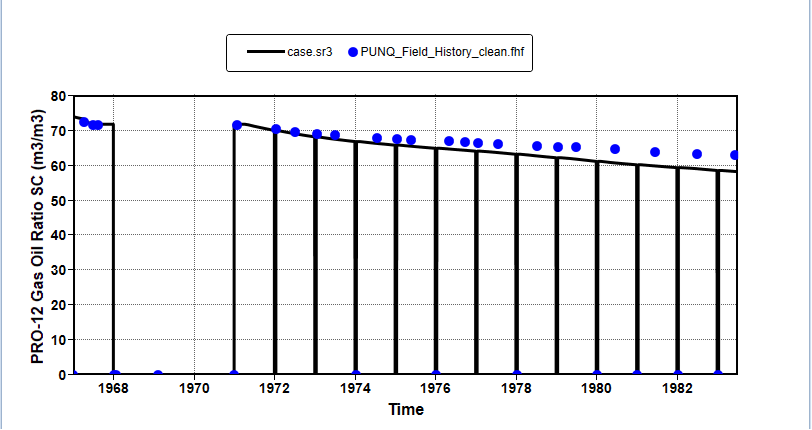}
        \\
        \small (a) PRO-1 gas--oil ratio (GOR).
        &
        \small (b) PRO-12 gas--oil ratio (GOR).
        \\[2mm]
        \includegraphics[
            width=0.315\textwidth,
            height=0.205\textheight,
            keepaspectratio
        ]{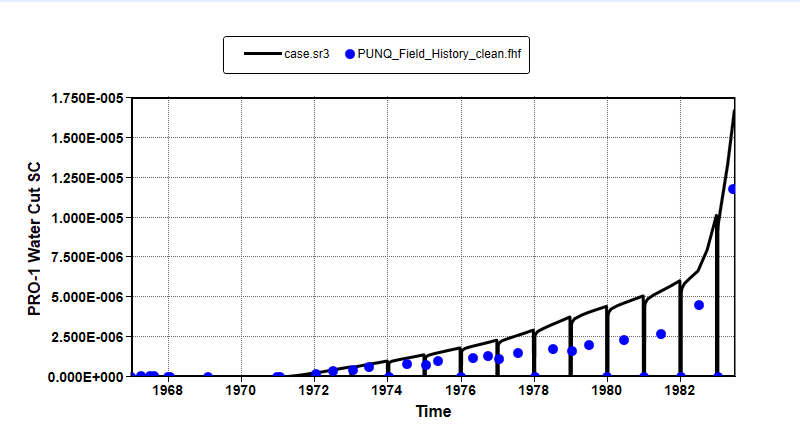}
        &
        \includegraphics[
            width=0.315\textwidth,
            height=0.205\textheight,
            keepaspectratio
        ]{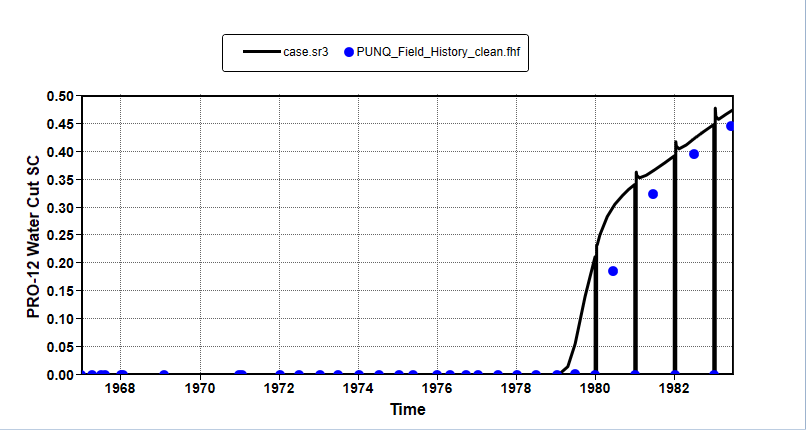}
        \\
        \small (c) PRO-1 water cut (WC).
        &
        \small (d) PRO-12 water cut (WC).
    \end{tabular}
    \caption{Simulated and observed GOR and water-cut trajectories for PRO-1 and PRO-12. Black curves show the matched reservoir simulation, and blue markers show the field-history observations.}
    \label{fig:gor_wc_pro1_pro12}
\end{figure*}

\begin{figure*}[t]
    \centering
    \setlength{\tabcolsep}{2pt}
    \renewcommand{\arraystretch}{1.0}

    \begin{tabular}{ccc}
        \includegraphics[
            width=0.30\textwidth,
            height=0.19\textheight,
            keepaspectratio
        ]{img/GOR_PRO1.png}
        &
        \includegraphics[
            width=0.30\textwidth,
            height=0.19\textheight,
            keepaspectratio
        ]{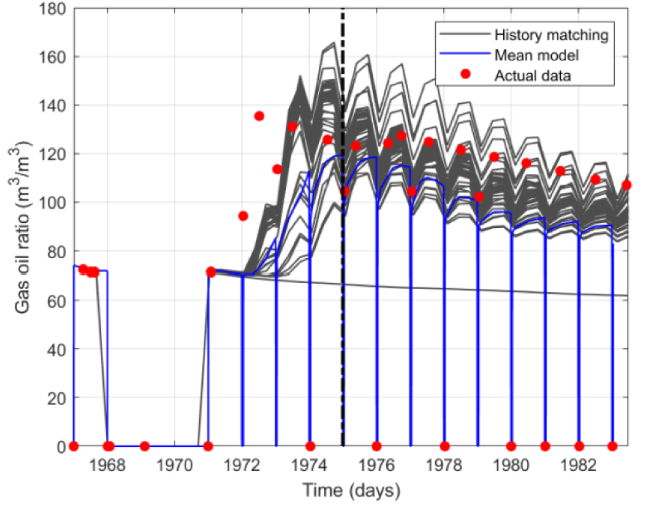}
        &
        \includegraphics[
            width=0.30\textwidth,
            height=0.19\textheight,
            keepaspectratio
        ]{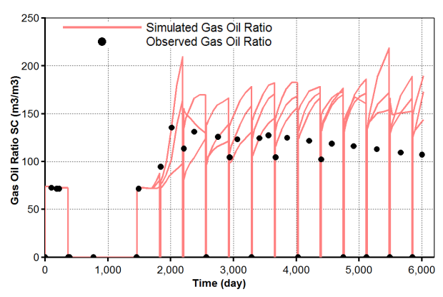}
        \\

        \small (a) RL-SAC: PRO-1
        &
        \small (b) ConvBiGRU-VAE+ES-MDA: PRO-1
        &
        \small (c) GP-VARS: PRO-1
        \\[2mm]

        \includegraphics[
            width=0.30\textwidth,
            height=0.19\textheight,
            keepaspectratio
        ]{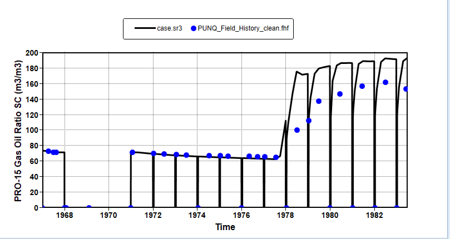}
        &
        \includegraphics[
            width=0.30\textwidth,
            height=0.19\textheight,
            keepaspectratio
        ]{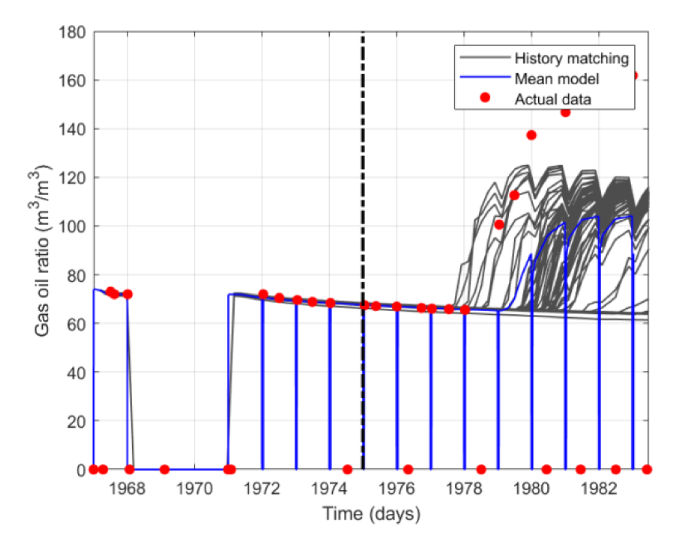}
        &
        \includegraphics[
            width=0.30\textwidth,
            height=0.19\textheight,
            keepaspectratio
        ]{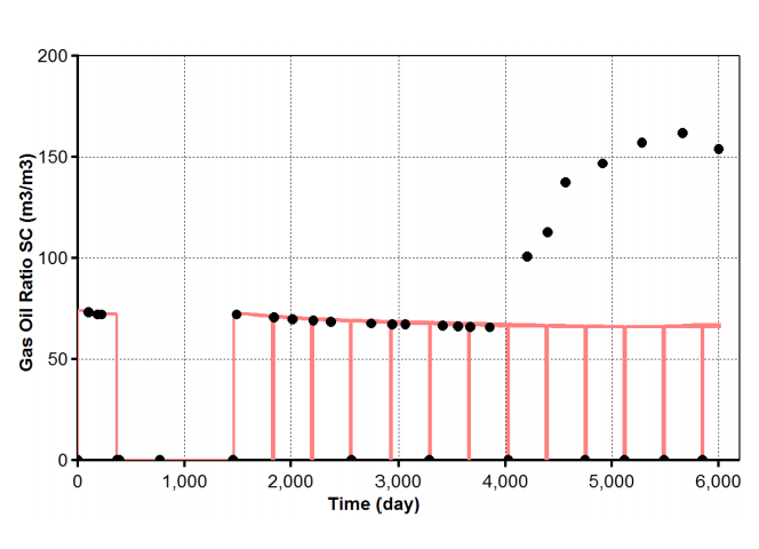}
        \\

        \small (d) RL-SAC: PRO-15
        &
        \small (e) ConvBiGRU-VAE+ES-MDA: PRO-15
        &
        \small (f) GP-VARS: PRO-15
    \end{tabular}

    \caption{GOR comparison for PRO-1 and PRO-15 on PUNQ-S3. Panels (a,d) show RL-SAC; panels (b,e) reproduce ConvBiGRU-VAE+ES-MDA~\cite{alguliyev2022history}; and panels (c,f) reproduce GP-VARS~\cite{rana2018gpvars}. Axis ranges follow the original sources.}
    \label{fig:gor_baseline_comparison}
\end{figure*}

\subsubsection{Production-Trajectory Matching}

Figure~\ref{fig:gor_wc_pro1_pro12} provides trajectory-level comparisons of GOR and water cut for PRO-1 and PRO-12. Together with the well-response metrics, these curves show how the aggregate match translates to individual production responses. The largest deviations in the full well-level evaluation occur in the water responses of PRO-1 and PRO-15, consistent with Table~\ref{tab:well_category_results}. This response-specific analysis prevents the strong aggregate scores from obscuring localized trajectory mismatch.

\subsection{Comparison with Existing Baselines}

Direct comparison with published PUNQ-S3 results is difficult because studies vary in uncertain parameters, historical intervals, objectives, and metrics~\cite{liu2023automatic,alguliyev2022history,rana2018gpvars}. The available EnKF and ConvBiGRU-VAE+ES-MDA study reports GOR rather than raw gas rate. We therefore restrict the numerical context to oil rate and BHP, whose response definitions are consistent across the compared studies.

On the four common wells, RL-SAC obtains a mean oil-rate $R^2$ of $0.9246$, compared with $0.9224$ for EnKF and $0.9401$ for ConvBiGRU-VAE+ES-MDA. For BHP, the corresponding values are $0.8452$, $0.8546$, and $0.9342$, respectively~\cite{alguliyev2022history}. Table~\ref{tab:contextual_comparison} presents these values as contextual evidence rather than a controlled ranking.

\begin{table}[t]
    \centering
    \caption{Contextual $R^2$ comparison on PRO-1, PRO-11, PRO-12, and PRO-15. Protocols differ across methods.}
    \label{tab:contextual_comparison}
    \small
    \begin{tabular}{@{}lccc@{}}
        \hline
        Response & RL-SAC & EnKF & DL+ES-MDA \\
        \hline
        Oil rate & 0.9246 & 0.9224 & 0.9401 \\
        BHP      & 0.8452 & 0.8546 & 0.9342 \\
        \hline
    \end{tabular}
\end{table}

The remaining baselines provide complementary computational evidence. GP-VARS was evaluated on a 45-parameter PUNQ-S3 problem and reportedly required approximately four times fewer numerical simulations than Differential Evolution~\cite{rana2018gpvars}. The FNO study reports low pressure and saturation prediction errors and a reduction in training simulations through data augmentation~\cite{badawi2025neural}. These values describe sample efficiency or forward-surrogate fidelity, not directly comparable PUNQ-S3 history-matching accuracy.

A fair efficiency comparison must count all full-physics evaluations, including surrogate-training simulations and every ensemble member~\cite{badawi2025neural,ma2024fno,abdulkareem2026e2co,aanonsen2009enkf,emerick2013esmda}. Because compatible traces are unavailable, a controlled ranking would require identical objectives, parameterizations, budgets, and random-seed protocols.

The numerical comparison excludes GOR because the available studies use different reporting intervals and evaluation protocols. Figure~\ref{fig:gor_baseline_comparison} therefore provides a complementary trajectory-level comparison for PRO-1 and PRO-15. These wells represent distinct behaviors: PRO-1 contains a cyclic high-GOR regime, whereas PRO-15 exhibits a delayed late-time increase. The comparison emphasizes transition timing, long-term trend, cyclic structure, and agreement with observations rather than pixel-level distances or a shared error metric.

For PRO-1, all three methods identify the transition from the initial low-GOR regime to the cyclic high-GOR regime. RL-SAC reproduces the repeated shut-in periods and the decreasing envelope of the post-transition peaks, although it moderately overestimates their amplitudes. ConvBiGRU-VAE+ES-MDA closely follows the assimilation-period observations but underestimates several later points. GP-VARS captures the transition but exhibits larger oscillations and overestimates several peak values.

The distinction is more pronounced for PRO-15. The observed GOR remains approximately stable during the early history before increasing sharply after approximately 4,000 days. GP-VARS largely retains the earlier level and does not reproduce this late-time transition. ConvBiGRU-VAE+ES-MDA captures the onset of the increase, but its posterior mean remains below several late observations. RL-SAC recovers the transition and subsequent increasing regime, although it overestimates the late-time magnitude.

Across both wells, RL-SAC shows strong structural fidelity but retains amplitude bias. The qualitative evidence therefore supports the quantitative results without establishing universal superiority. A controlled comparison would require the original trajectories, identical time alignment, a shared GOR objective, and the same full-physics simulation budget. Figure~\ref{fig:gor_baseline_comparison} should consequently be interpreted together with Tables~\ref{tab:category_results}, \ref{tab:well_category_results}, and \ref{tab:contextual_comparison}.

\section{Discussion}
\label{sec:discussion}

From an AI perspective, SAC acts as an adaptive stochastic proposal distribution over simulator inputs. Replay reuses expensive IMEX interactions across updates, while entropy regularization discourages premature collapse to a narrow region. The final aggregate NMSE of $0.0285$ and $R^2$ of $0.9324$ show that this policy-search interface can identify a strong full-physics calibration in a compact continuous space.

The experiment does not establish general SAC superiority. The environment is bandit-like, the policy is trained on one reservoir with one seed, and published baselines do not share the same objective or simulator budget. Thus, the evidence supports feasibility rather than long-horizon credit assignment, cross-reservoir generalization, or a definitive sample-efficiency advantage. Controlled multi-seed comparisons with equal-budget random search and derivative-free optimizers remain necessary.

The weak well-level water-cut NMSE for PRO-1 and PRO-15 is primarily associated with the very small magnitude of their observed responses, which are on the order of $10^{-4}$--$10^{-5}$. At this scale, even small absolute simulation errors become large relative errors after normalization, and accurately reproducing minute changes in water production is more difficult for the full-physics simulator. Thus, the weak normalized scores do not indicate a comparably large field-scale water imbalance. Nevertheless, aggregate metrics should be paired with response-level diagnostics, and a robust reward could combine the category-level loss with a worst-well penalty, for example,
\begin{equation}
    \mathcal{L}_{\mathrm{robust}}
    =
    \frac{1}{|\mathcal{W}||\mathcal{C}|}
    \sum_{w\in\mathcal{W}}
    \sum_{c\in\mathcal{C}}
    \mathcal{L}_{w,c}
    +
    \lambda_{\max}
    \max_{w,c}\mathcal{L}_{w,c},
\end{equation}
where $\mathcal{W}$ and $\mathcal{C}$ denote wells and response categories. This term would discourage sacrificing a difficult low-volume response to improve aggregate reward.

Future evaluation should use equal-budget, multi-seed ablations of entropy, replay, uniform proposals, and the physics penalty. Scaling beyond five global parameters will require structured spatial or latent actions and more principled constraint handling, such as primal--dual updates. Reporting failed runs, wall-clock cost, and best-so-far mismatch versus full-physics calls would also support fairer comparisons with other optimizers.

\section{Conclusions}

We presented a physics-constrained SAC workflow for black-box reservoir calibration. The policy proposes continuous reservoir parameters, learns from replayed simulator interactions, and is evaluated directly by CMG IMEX. On PUNQ-S3, the best candidate reaches category-macro NMSE $0.0285$, $R^2=0.9324$, and $C_{\mathrm{match}}=97.23\%$, while well-level analysis reveals localized water-rate failures hidden by pooled scores.

The main methodological contribution is the separation of learned proposal generation, physical constraint feedback, and simulator-grounded model selection. This decomposition permits future agents or optimizers to be substituted without changing the authoritative evaluation pipeline, and it provides a clear accounting unit: one candidate corresponds to one full-physics simulator call.

The same interface applies to other scientific calibration problems with continuous parameters and costly simulators. It requires a reproducible case generator, output parser, task-specific mismatch, and validity checks, while domain simulators remain responsible for scientific validity.

These results are a single-benchmark proof of concept, not evidence of universal superiority. Future work should use equal-budget baselines, multiple seeds, spatial parameterizations, robust well-aware rewards, and cross-reservoir evaluation, with final model selection retaining full-physics validation.

\begingroup
\hbadness=10000
\bibliography{aaai2026}
\endgroup

\end{document}